\documentclass{astlb} 

\usepackage[hyperfootnotes=false]{hyperref}
\usepackage[utf8]{inputenc}
\usepackage[T2A]{fontenc}
\usepackage{color}
\usepackage{epsf}
\usepackage{amsmath}
\usepackage{amsfonts}
\usepackage{amssymb}
\usepackage{epsf}
\usepackage{graphicx}
\usepackage{gensymb}
\usepackage{float}
\usepackage{changes}
\usepackage{aas_macros}

\newcommand{\art}{ART-XC}

\begin{document}

\journalinfo{2023}{49}{5}{0}[0]

\title{Spectral and temporal analysis of the Supergiant Fast X-ray Transient IGR J16195-4945 with SRG/ART-XC}
%\title{Спектральный и временной анализ быстрого рентгеновского транзиента IGR J16195-4945 по данным наблюдений СРГ/ART-XC}

\author{
  M.~N.~Satybaldiev\email{maksatsatybaldiev@gmail.com}\address{1,2}, I.~A.~Mereminskiy\address{1},
  A.~A.~Lutovinov\address{1}, D.~I.~Karasev\address{1}, A.~N.~Semena\address{1}, A.~E.~Shtykovsky\address{1}
    \addresstext{1}{Space Research Institute, Moscow}
    \addresstext{2}{Moscow Institute of Physics and Technology}
}

\shortauthor{Satybaldiev et al.}
\shorttitle{IGR J16195-4945}

\submitted{28.03.2023}
\revised{19.05.2023}
\accepted{03.06.2023}

\begin{abstract}
We present the results of the analysis of the SRG/ART-XC observation of the Supergiant Fast X-ray Transient IGR J16195-4545 performed on March 3, 2021. 
Six bright flares are present in the light curve, with no significant change in hardness occuring during these flares. The spectrum is described with an absorbed power law model with a high energy exponential cutoff showing heavy absorption, with $N_H=(12\pm2)\times 10^{22}\text{ cm}^{-2}$ and $\Gamma=0.56\pm 0.15$, $E_{cut}=13\pm 2$ keV. Adopting the Bayesian block decomposition of the light curve, we measured the properties of the observed flares (duration, rise time, waiting time, released energy and pre-flare luminosity), which are consistent with the quasi-spherical subsonic accretion model. The stellar wind velocity of the supergiant is estimated to be  $v_{w} \approx 500$ km s$^{-1}$. Additionally, the system was found to have an unusual near-IR variability.
  \\
\end{abstract}

\section*{Introduction}

Supergiant Fast X-ray Transients (SFXT) are a sub-class (\citealt{sguera05}; \citealt{negueruela06};  \citealt{smith06}; \citealt{zand04}; \citealt{grebenev10}; \citealt{sidoli17}) of High Mass X-ray Binaries (HMXB) in which the inhomogeneous stellar wind from a blue supergiant is accreted onto a compact object (neutron star or black hole). 
A distinctive feature of such systems is their X-ray variability. 
They demonstrate short sporadic flares lasting for $\sim 10^3 - 10^4$ s, during which the X-ray luminosity exceeds $10^{35}$ erg s$^{-1}$ and, in some cases, even reaches $10^{37}$ erg s$^{-1}$, while between the flares the X-ray luminosity is  $10^{32} - 10^{34}$ erg s$^{-1}$.  

Many different models were proposed to explain such varibility: the accretion of dense wind clumps \citep{zand05, walter07},  centrifugal and/or magnetic gating (\citealp{grebenev07}; \citealp{bozzo08}), accretion of an asymmetric stellar wind in systems with wide and eccentric orbits\citep{sidoli07},  quasi-spherical subsonic settling accretion \citep{shakura14}.

The transient X-ray source IGR J16195-4945 was discovered by the IBIS/ISGRI\citep{lebrun03} telescope onboard the INTEGRAL observatory (\citealt{winkler03}; \citealt{kuulkers21}).
On September 26, 2003,  the source exhibited a bright flare lasting $\sim 1.5$ hours with an average flux of $\sim 35$ mCrab in the 20-40 keV range\citep{sguera06}.
Such flaring activity made IGR J16195-4945 a SFXT candidate.
The soft X-ray counterpart of this source was identified in archival observations of ASCA \citep{sidoli05}.

The {\it Chandra} observation, performed on April 29 2005, allowed for the refinement of the sources position to RA = 16h 19m 32.20s, Dec= $-49\degree$ 44' 30.7'' (J2000),  with an accuracy of 0.6'' \citep{tomsick06}.
It made possible to identify the source counterparts in the near- and mid-infrared catalogs - 2MASS (2MASS J16193220-4944305) and GLIMPSE (G333.5571 + 00.3390), respectively. Near-IR spectroscopy determined that the donor star is an ON9.7Iab blue supergiant star \citep{coleiro13}.

\citealt{morris2009} reported on results of {\it Suzaku} observation, performed in September 2006. They observed a bright flare with a duration of $\sim 5000$ s and a peak flux $\sim 10\times$ brighter than the prior emission level, that once again confirmed that IGR J16195-4945 belonged to the class of SFXTs. The subsequent spectral analysis  revealed that the system is heavily absorbed ($N_H \simeq 1.1\times10^{23}\text{ cm}^{-2}$), with no significant iron lines in the spectrum (EW$<43$ eV). 

Using the Swift/BAT survey data obtained between December, 2004 and March 2015 and all the available Swift/XRT pointed observations, \citet{cusumano16} demonstrated that IGR J16195-4945 is an eclipsing binary with an orbital period of $3.945$ days. The eclipse lasts for $\sim 3.5 \%$ of the orbital period. 

In this paper, we report the results of the temporal and spectral analysis of the long continuous observation of IGR J16195-4945 by the Mikhail Pavlinsky ART-XC telescope. The observed flaring activity was interpreted in terms of the subsonic settling accretion model. Moreover, the object was found to have an unusual near-IR variability with an amplitude of 0.1 magnitude in K-filter. 

\section{Observations}
IGR J16195-4945 was observed by the Mikhail Pavlinsky ART-XC telescope \citep{pavlinsky2021} on March 3, 2021 for a total exposure of 86 ks. 
The ART-XC data was processed using the {\sc artproducts} v1.0 software with the latest calibration files {\sc v20220908}. 
Spectra and light curves were extracted from a circular region  of radius $R = 135\arcsec$ centered at the source position.
For the extraction of the light curve, a wide energy range of 4-20 keV was used, along with two sub-ranges: 4-8 keV and 8-20 keV. Spectral analysis, on the other hand, was performed in the energy range of 5-25 keV. 
Due to the instrument's response requiring more precise calibration at energies below 5 keV, this energy region was excluded from the spectral analysis.

\section{Temporal analysis}
\label{sec:timing}

\begin{figure*}
\centering
   \includegraphics[width=0.95\textwidth]{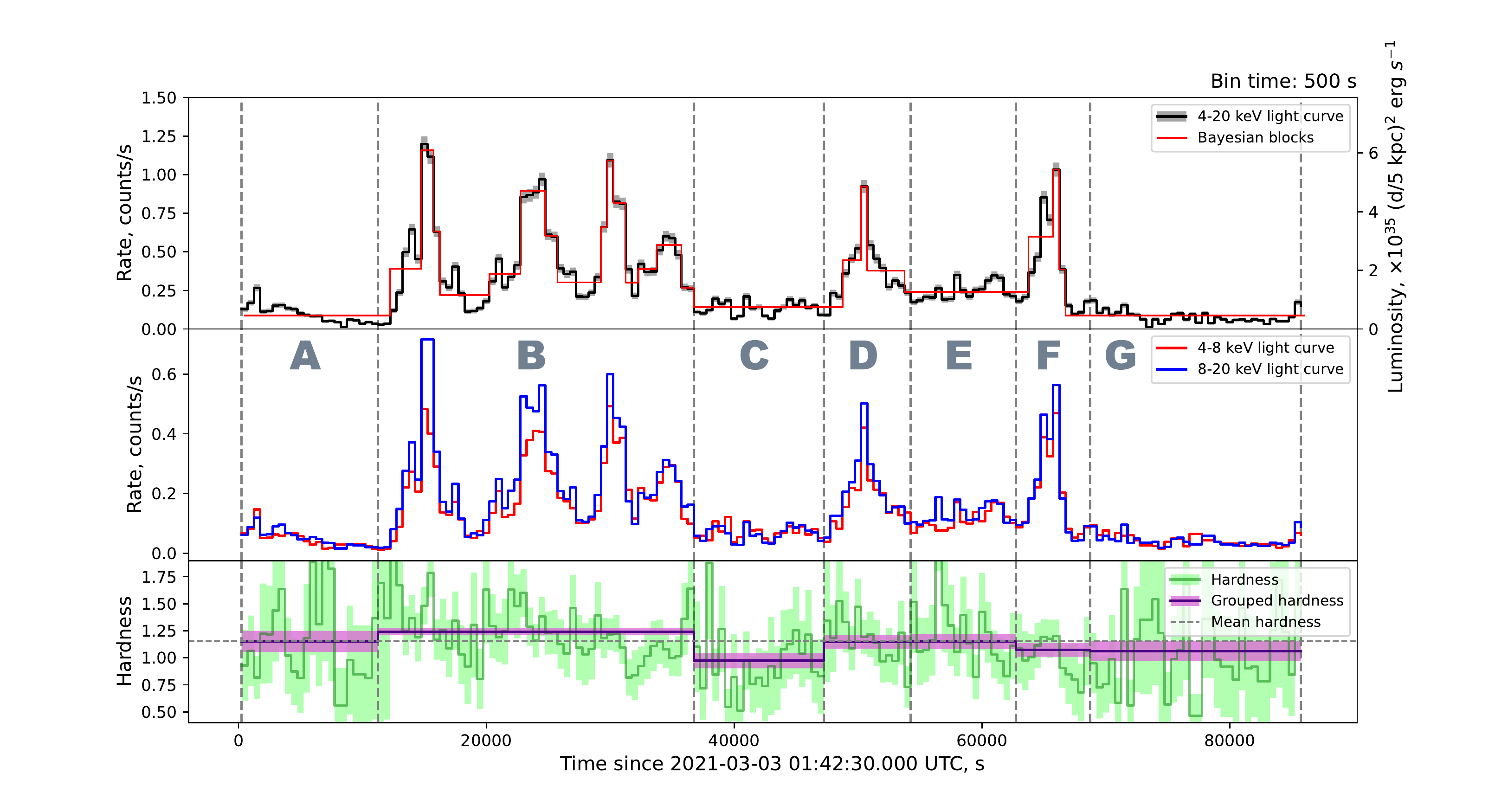}
     \caption{{\it Top panel:} 500 s binned light curve (4-20 keV) and its Bayesian block segmentation (division into intervals with a statistically significant difference in the count rate).\\
     {\it Middle panel:} light curves in soft (4-8 keV) and hard (8-20 keV) energy ranges.\\
     {\it Bottom panel:} full and segment-averaged hardness (the ratio of count rates in the hard and soft energy ranges).
     }
     \label{fig:hardness}
\end{figure*}

Figure \ref{fig:hardness} shows the light curve of the source with 500s bin covering a broad energy band of 4-20 keV.
The light curve clearly shows flares, which are characteristic of SFXTs and have durations of a few thousand seconds.
For further analysis we have divided the lightcurve into several segments based on the source's activity state: quiescent (segments A, C, G), intermediate (E), and active (B, D, F). 
Unfortunately, due to the uncertainty in the ephemeris of the binary system, it was not possible to determine the exact orbital phases corresponding to the observation.
However, the expected eclipse, with a duration of 12 ks \citealt{cusumano16}, did not occur during the observation. The dynamic range, defined as the ratio of the count rate at the peak of the brightest flares to the average rate in the quiescent state, is$\sim 11$. 

We have also extracted light curves in the soft (4-8 keV) and in hard (8-20) energy ranges. 
In SFXTs, where the neutron star is surrounded by a clumpy wind from its companion, significant variations in observed absorption can be expected when the wind clumps pass between the NS and the observer \citep{oskinova12}, significantly changing the observed flux in the soft X-rays.
However, the soft and hard light curves were found to be similar, with no apparent hardening episodes.
% The hardness averaged over segments was estimated (the ratio of count rates in the hard and soft energy ranges). 
For a more detailed study, we have also calculated the hardnesses  (the ratio of count rates in the hard and soft energy ranges), averaged over the segments of time of different activity states. It is noticeable that the light curve becomes slightly harder during flares, but in general it could be concluded that the source demonstrates “colorless” variability known in other SFXTs \citep[see e.g.][]{sidoli20}.

In some SFXTs, periodic flux modulations resulting from the rotation of the NS have been observed, typically with periods of 10-1000 s \citep{walter15}.
Using the unbinned events we searched for the presence of periodicities in light curve in the range of 10-1000 s using the epoch folding method \citep{leahy83}.
No significant periodicities were detected for periods up to approximately $\sim 300$ s at a significance level of $4.5\sigma$.
In the periodogram, a signal is observed for periods of $300-1000$ s. However, this signal is not associated with the rotation of the NS but rather corresponds to the observed flares, which have durations of $\sim 1000$ s.

\begin{table*}[h!]

\vspace{6mm}
\centering
\caption{Parameters of the best-fit for the spectra of IGR J16195-4945}
% {const*tbabs*cflux*cutoffpl}
% \small
\footnotesize
\vspace{5mm}\begin{tabular}{c c c c c c} \hline
% Parameter & Best fit value & Units \\
Segment & $N_H$, cm$^{-2}$ & $\Gamma$ & $E_{cut}$, keV & $\chi^2$ / d.o.f. & F[4-20 keV], $\text{ erg}\text{ s}^{-1} \text{ cm}^{-2}$\\
\hline\hline
Full ART-XC + XRT + BAT & $(12\pm2)\times10^{22}$  & $0.56\pm0.15$ & $13\pm2$ & 231.29 / 185 & $(2.5\pm 0.1)\times10^{-11}$ \\
ART-XC & $31\pm 15$ & $1.09\pm 0.41$ & $19^{+28}_{-7}$ & 191.15 / 157 & $(2.9^{+0.4}_{-0.2})\times 10^{-11}$\\
Full ART-XC (fixed $N_H$)& $12\times10^{22}$ & $0.67\pm 0.27$ & $15^{+12}_{-5}$ & 196.49 / 158 & $(2.4\pm 0.3)\times 10^{-11}$\\
ART-XC: ACG (fixed $N_H$)& $12\times10^{22}$  & $0.58^{+0.84}_{-0.97}$ & $10^{+88}_{-5}$ & 187.16 / 158 & $(0.9\pm 0.1)\times 10^{-11}$\\
ART-XC: BDF (fixed $N_H$) & $12\times10^{22}$ & $0.59\pm0.27$ & $15^{+9}_{-5}$ & 184.15 / 158 & $(4.1\pm 0.2)\times10^{-11}$\\
 \hline
\end{tabular}
\label{table:fit}
\end{table*}

\begin{figure}[h!]
\centering
      \includegraphics[width=0.5\textwidth]{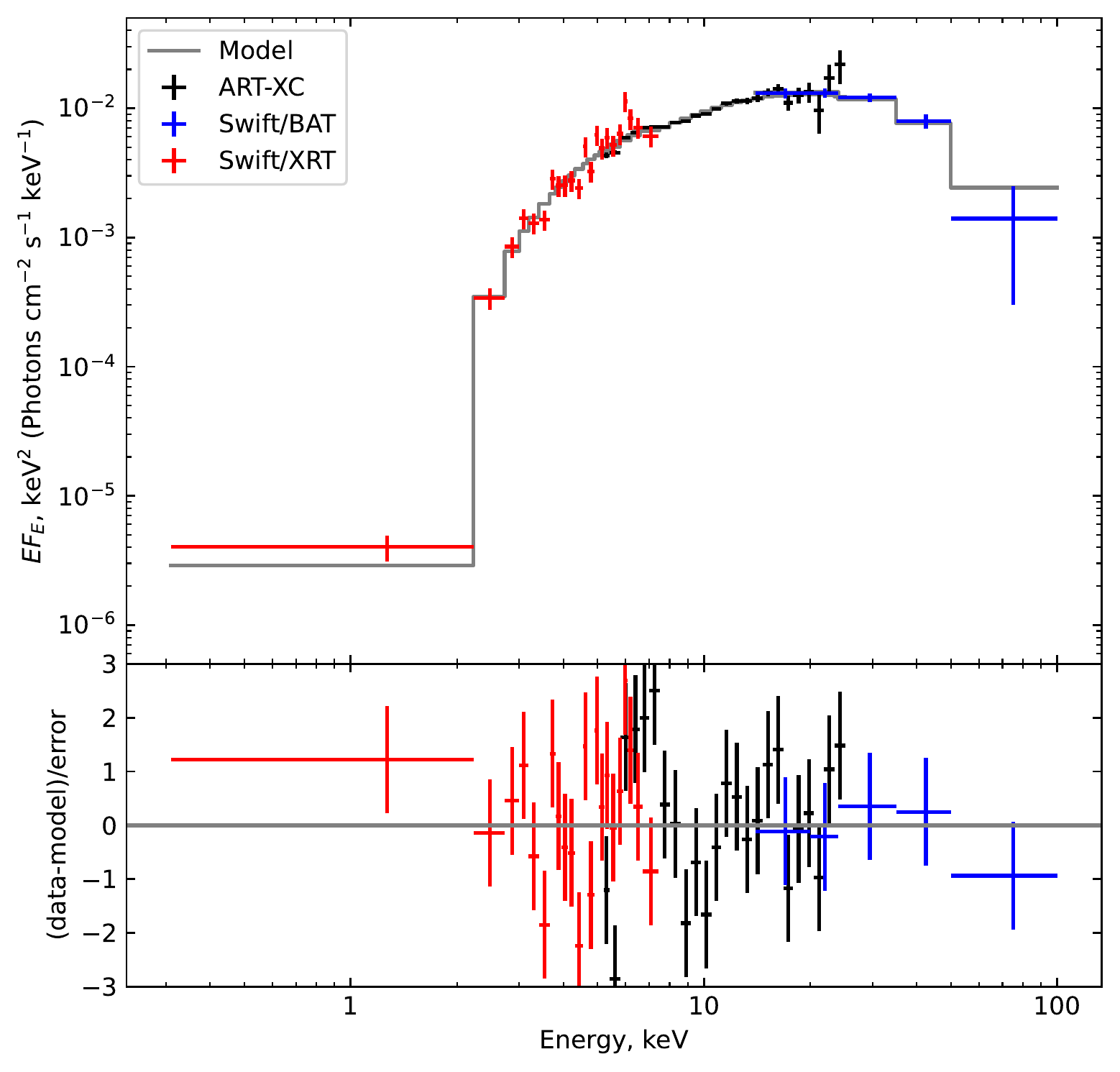}
     \caption{X-ray spectrum of IGR J16195-4945 measured by ART-XC (black), Swift/XRT (red) and Swift/BAT (blue) telescopes. Best-fit model is shown with solid grey line.}
     \label{fig:spectrum}
\end{figure}

\begin{table*}[h!]

\vspace{6mm}
\centering
\caption{Measured flare properties}
\footnotesize
\vspace{5mm}\begin{tabular}{c c c c c c c c} \hline
$\#$ & Energy, $10^{38}\text{ erg (1-10 keV)}$& Waiting time, s & Duration, s & Rise time, s &  Pre-flare $L_x$, $10^{34}\text{ erg s}^{-1}$ (1-10 keV)\\ 
 \hline\hline
1 & $4.4\pm0.2$ & 15000(*) & 4000 & 2500  & $1.6\pm0.1$\\
2 & $5.9\pm0.3$ & 8500 & 5500 & 2500 &  $4.0\pm0.2$\\
3 & $3.1\pm0.3$ & 6250 & 2000 & 500  & $5.5\pm0.3$\\
4 & $3.5\pm0.2$ & 4750 & 4500 & 1500  & $5.4\pm0.4$\\
5 & $4.1\pm0.2$ & 15750 & 5000 & 1500  & $2.6\pm0.1$\\
6 & $3.4\pm0.2$ & 15500 & 3000 & 2000  & $4.3\pm0.2$\\
\hline
\end{tabular}

\label{table:flares}
\end{table*}

\section{Spectral Analysis}

The long continuous observation performed by the \art\, made it possible to obtain a good spectrum covering 5-25 keV energy range. 
To describe the spectrum we have chosen an absorbed power law with high energy exponential cutoff model (\texttt{tbabs*cutoffpl}) that is commonly used to describe SFXT spectra \citep[see e.g.][]{romano15}.
Spectra from all seven telescope modules of \art\, were fitted simultaneously with \texttt{XSPEC v12.12.1} package. 
All errors are given in 90\% confidence interval. The value and error of the unabsorbed flux were estimated using the \texttt{cflux}. 

The measured parameters - photon index and cutoff energy - were found to be in close agreement with the values obtained from the joint spectrum of Swift/XRT and Swift/BAT \citep{cusumano16}.
However, due to the limited coverage in the low-energies region, an accurate measurement of absorption from the \art\ data alone is not feasible. The obtained parameters are presented in Table \ref{table:fit}. 

To build a broadband spectrum, we added to the \art\, data the 105-month averaged Swift/BAT spectrum \citep{oh18} and the averaged Swift/XRT spectra of observations where count rate was higher than $5\times10^{-2}$ counts/sec (observations 1, 2, 4, 5, 21, 22 from Table 1 of \citealt{cusumano16}). 
The resulting spectrum and residuals are shown in Figure~\ref{fig:spectrum}. 
The expansion of the energy range made it possible to accurately measure the absorption $N_{H} = (12\pm2)\times10^{22}$ cm$^{-2}$ and cutoff energy $13\pm2$ keV. 
Cross-normalization coefficients were found to be $1.991$ and $1.073$ for XRT and BAT, respectively.

To convert the obtained light curve from counts/s to physical fluxes, we fitted the \art\ spectrum with a fixed $N_{H} = (12\pm2)\times10^{22}$ cm$^{-2}$. 
This allowed us to determine a conversion factor of $K_{bol}=5.3\times 10^{35}$ erg counts $^{-1}$ which relates the observed count rate to the total unabsorbed luminosity of the source in the energy range of 0.1-100 keV, assuming the distance to the system is 5 kpc \citep{tomsick06}. 
We further used this factor to plot the bolometric luminosity curve (Fig.~\ref{fig:hardness}), assuming that the most of the energy release in the system occurs in the X-rays.
Furthermore, we divided the spectra into two groups: active states (intervals B, D, F) and "low" states (intervals A, C, G), and separately fitted them using the same model.
Despite the average fluxes differing by a factor of 4 between these states, no significant differences in the spectral parameters were observed. 
This finding confirms the "colorless" nature of the observed variability.

\begin{figure*}[h]
\centering
   \includegraphics[width=0.8\textwidth]{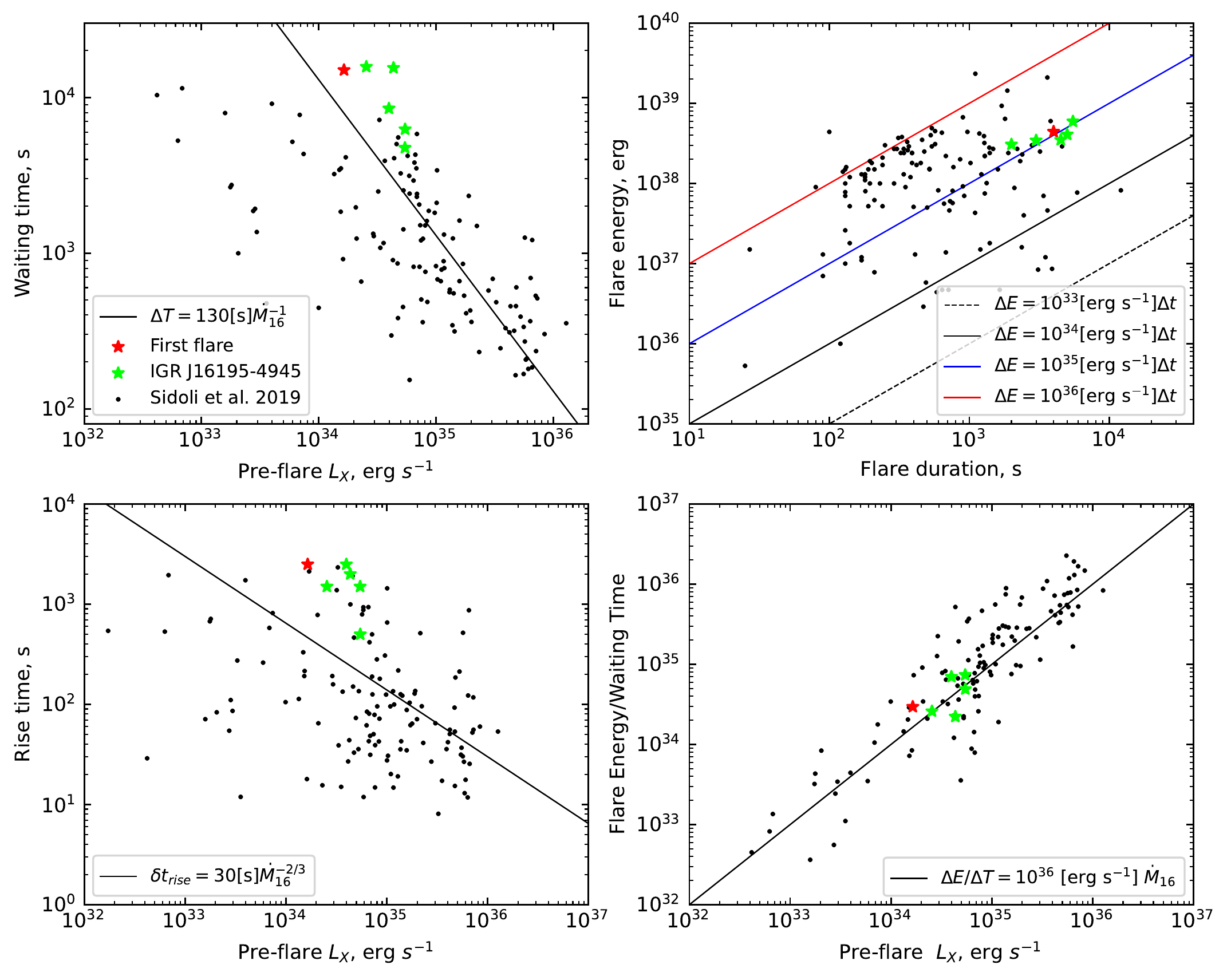}
     \caption{Flare properties of IGR J16195-4945 and dependencies described by the "settling" accretion model}
     \label{fig:flares_characteristics}
\end{figure*}

\section{Flare properties}

The average bolometric luminosity of IGR J16195-4945 during the observation was $L_{mean}=(1.38\pm0.05)\times 10^{35} (d/5\text{ kpc})^2\text{ erg s}^{-1}$.
Thus, for the range of possible distances to the system (5--15 kpc, \citealt{tomsick06}) the average luminosity turns out to be less than the critical value $L_{crit}\approx 4\times10^{36}$ erg s$^{-1}$ \citep{shakura15}. Therefore, in this system matter from supergiant wind could be accreted through "settling" subsonic accretion. 

\citealt{sidoli19} discussed the occurrence of flares resulting from plasma portions from a hot shell penetrating the magnetosphere by the Rayleigh-Taylor Instability (RTI). 
A series of such flares can arise due to fluctuations in the density or velocity of the stellar wind, for example, due to the arrival of a wind clump.

For such flares, dependencies were derived between various parameters, such as such as flare energy, waiting time, pre-flare luminosity and others. These relations have the form:
flare waiting time $\Delta T$ on pre-flare luminosity ($L_{\rm X, pre}$, pre-flare):
$$\Delta T \approx 130 \text{[s]}\bigg(\cfrac{\alpha}{0.03}\bigg)A\zeta^{2/9}\mu_{30}^{2/3}\dot{M}^{-1}_{16},$$
energy released in a flare $\Delta E$ on its duration $\Delta t$:
$$\Delta E\approx 3\times 10^{35}\text{[erg s}^{-1}\text{]}\bigg(\cfrac{\alpha}{0.03}\bigg)A\zeta^{2/9}\mu_{30}^{2/3}v_8^{3}\Delta t,$$
flare rise time $\delta t_{rise}$ on $L_{\rm X, pre}$:
$$\delta t_{rise}\simeq 30\text{[с]}\zeta^{4/27}\mu_{30}^{7/9}\dot{M}_{16}^{-2/3},$$
ratio between the flare energy and the waiting time $\Delta E/\Delta T$:
$$\cfrac{\Delta E}{\Delta T}=10^{36}\text{[erg s}^{-1}\text{]} \dot{M}_{16},$$
where the mass accretion rate $\dot{M}_X=10^{16}\text{[g s}^{-1}\text{]}\dot{M}_{16}$
related to pre-flare luminosity as $L_{\rm X, pre}=0.1\dot{M}_Xc^2$, $\alpha\sim 0.03$ - , the dimensionless factor, defining the non-linear growth rate , $A \lesssim 1$ is the effective Atwood number, $\zeta\lesssim 1$  characterizes the size of the RTI region in units of the magnetospheric radius $R_m$, 
$\mu=10^{30}\text{[G cm}^{3}\text{]}\mu_{30}$ -  the NS magnetic moment, $v=10^8\text{[cm s}^{-1}\text{]}v_8$ - the relative wind velocity.   

These relations along with flare properties measured by \citealt{sidoli19} for 9 known SFXTs observed by XMM-{\it Newton} observatory are shown on Figure~\ref{fig:flares_characteristics}.

We have reproduced the analysis from \citealt{sidoli19} using a Bayesian block decomposition \citep{scargel13} of the light curve (see Figure ~\ref{fig:hardness}) and measured the properties of the observed flares that are given in Table ~\ref{table:flares}.
For the first flare, the waiting time was calculated relative to the start of observation.
To compare {IGR~J16195-4945} with other SFXTs the flare properties were recalculated to a softer energy range of 1-10 keV. 
For this, the average spectrum of the source for the entire observation was used, since it was previously shown that it does not significantly change with intensity.

In general, {IGR~J16195-4945} flares follow the predictions of settling accretion model.
However, it is necessary to recall that the exact distance to the system is not known and presented luminosities can be systematically underestimated by a factor up to $\approx10$, if the system is actually located farther, at a distance of 15 kpc. 
In the pre-flare $L_X$ versus waiting time plot,  {IGR~J16195-4945} flares lie above the $\Delta T \approx 130 \text{[s]} \dot{M}^{-1}_{16}$ line, that which may indicate a greater value of $(\frac{\alpha}{0.03})A\zeta^{2/9}\mu_{30}^{2/3}$.
The characteristic durations of the outbursts make it possible to estimate the velocity of the supergiant stellar wind.
From equation $\Delta t\approx 400\text{[s]}\bigg(\cfrac{v_w}{1000\text{[km s}^{-1}]}\bigg)^{-3}$ from \citealt{sidoli19}, it follows that for IGR J16195-4945 $v_{w} \approx 500 $ km s$^{-1}$, that is typical for HMXBs \citep{clumps17}

\section{Infrared variability}

{IGR~J16195-4945} is located in the Galactic plane ($l,b = 333.56\degree, 0.34\degree$), within the region of the sky covered by the VISTA Variables in the Via Lactea (VVV) near-infrared  survey \citep{minniti10VVV}.
From the VVV DR5 catalog we selected all reliable ($errBits\leq16$) $Ks$-filter measurements made inside 2'' aperture ({\it aperMag5}), in total 190 such measurements were performed during 2010-2015. We have also selected several neighboring stars ($40\arcsec$ and closer) for comparison. 

\begin{figure}[h]
\centering
   \includegraphics[width=0.95\columnwidth]{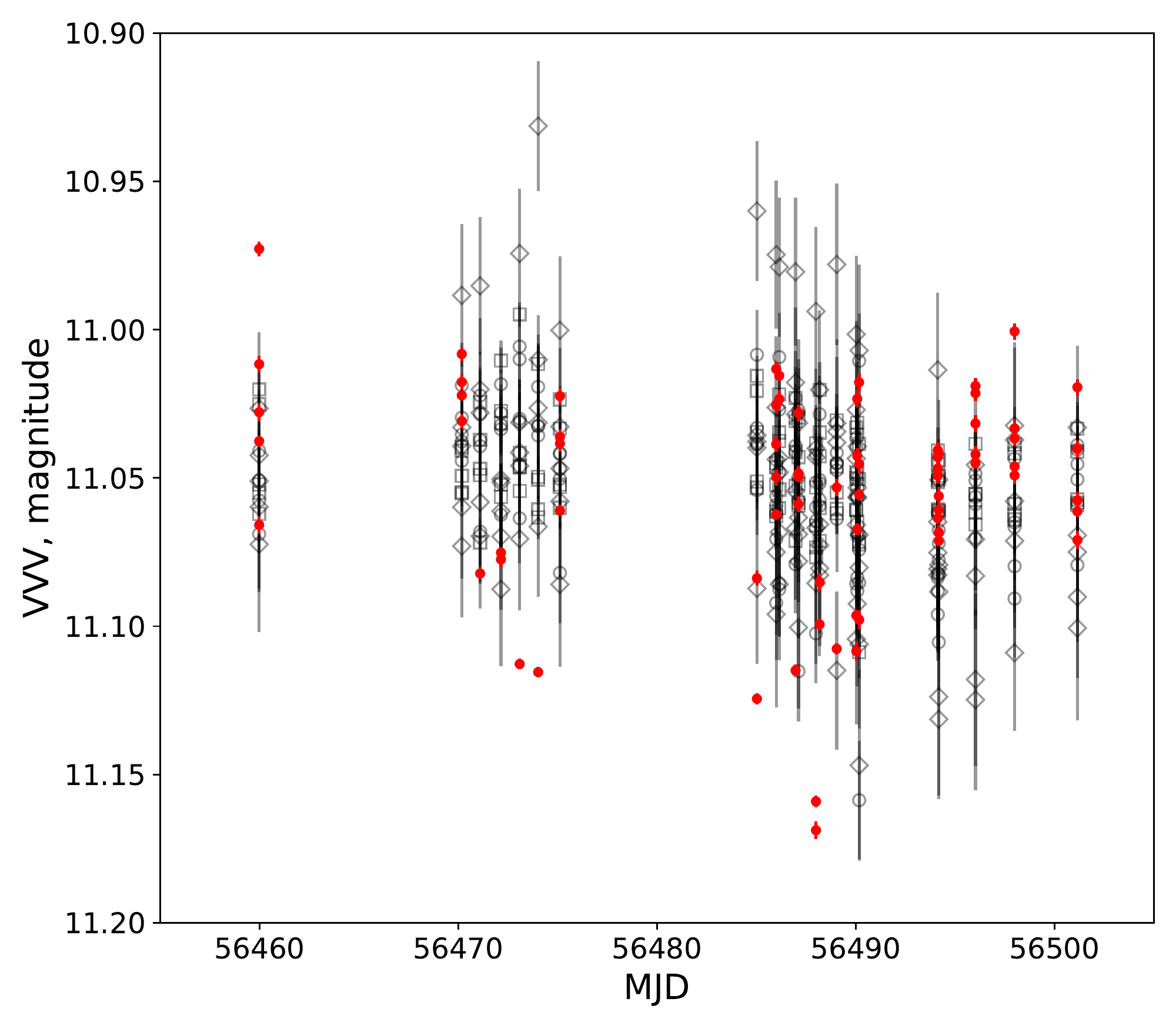}
   \caption{ {IGR~J16195-4945} in $Ks$ filter (red dots) according to the VVV survey. Black markers show magnitudes of comparison stars normalized to the average magnitude of {IGR J16195-4945}.}
   
     \label{fig:IR_lc}
\end{figure}

On the light curve (Figure \ref{fig:IR_lc}) it is clearly visible that the source exhibits rapid infrared variability, changing its brightness by 0.1-0.2 magnitudes over several days.
Such variability is atypical for single blue supergiants, as their typical variability amplitude in the optical range is usually within the range of $0.02...0.04$ magnitude \citep{buysschaert15,aerts17}.

Suchlike rapid changes in luminosity, by tens of percent in a few days, cannot be associated with the intrinsic variability of the supergiant star. Reprocessing of X-rays by the surface of the supergiant also cannot explain the observed rapid changes of its brightness, since even for the brightest observed flares the total bolometric luminosity is about 10$^{37}$ erg s$^{-1}$, which is an order of magnitude less that the bolometric luminosity of blue supergiants ($\gtrsim10^{5} $ L$_{\odot}$).
 
To determine the nature of the observed rapid variability, additional spectroscopic observations in the near-IR are required, preferably combined with X-ray monitoring of the system.

\section{Conclusion}
Thanks to the long-term continuous observation {IGR~J16195-4945} conducted by the Mikhail Pavlinsky ART-XC telescope of the SRG observatory in 2021, it was possible to study the variability of the source at scales of 10-10000 s. 
The light curve revealed the presence of six flares, each lasting several thousand seconds.
During flares, there was no strong change in the hardness of the X-ray radiation. 
No significant periodic signal was found.

The broadband spectrum constructed from the combined data of the ART-XC and the Swift observatory telescopes is well described by a power law model with an exponential cutoff at an energy of $\approx 13$ keV and a significant absorption $N_{H}\approx 10^{23}$ cm$^ {-2}$.

Since the average luminosity of the source turned out to be below the critical one, we compared the parameters of the observed flares in {IGR~J16195-4945} with flares in other SFXTs and with the predictions of the \citep{shakura12} "settling" accretion model.
The properties of flares from the source are close to those for other SFXT systems.
The system exhibits significant near-infrared rapid variability, which cannot be explained by simple assumptions and requires further investigation.
%----------------------------------------------------------------------------------------------

\acknowledgements
We thank the referees for their useful comments that helped to
improve this manuscript.  

This work is based on data from {\it Mikhail Pavlinsky} ART-XC X-ray telescope aboard the SRG observatory. The SRG observatory was built by the  Lavochkin Association (NPOL)  (part of State Corporation Roscosmos) in the interests of the Russian Academy of Sciences represented by its Space Research Institute (IKI) in the framework of the Russian Federal Space Program, with the participation of the Deutsches Zentrum für Luft- und Raumfahrt (DLR).
The \art\ team thank the Russian Space Agency, Russian Academy of Sciences and State Corporation Rosatom for the support of the SRG project and the Lavochkin Association (NPOL) with partners for the creation and operation of the SRG spacecraft (Navigator).

The work was supported by the Russian Science Foundation grant 19-29-11029.

\label{lastpage}

%%%%%%%%%%%%%%%%%%%%%%%%%%%%%%%%%%%%%%%%%%%%%%%%%%%%%%%%%%%%%%%%%%%%%%%%%%%%%%%%%%

\bibliographystyle{mnras}
\bibliography{engref}

\end{document}